\documentclass[conference]{IEEEtran}
\IEEEoverridecommandlockouts
\usepackage{cite}
\usepackage{amsmath,amssymb,amsfonts}
\usepackage{algorithmic}
\usepackage{graphicx}
\usepackage{textcomp}
\usepackage{subfig}
\usepackage{threeparttable}
\usepackage{xcolor}
\def\BibTeX{{\rm B\kern-.05em{\sc i\kern-.025em b}\kern-.08em
    T\kern-.1667em\lower.7ex\hbox{E}\kern-.125emX}}
\begin{document}

\title{FID: Function Modeling-based Data-Independent and Channel-Robust Physical-Layer Identification}


\author{Tianhang Zheng\textsuperscript{1}, Zhi Sun\textsuperscript{1}, Kui Ren\textsuperscript{1, 2}\\
	\textsuperscript{1} Department of Computer Science and Engineering, State University of New York at Buffalo\\ \textsuperscript{2} Institute of Cyberspace Research, Zhejiang University\\
    \{tzheng4, zhisun, kuiren\}@buffalo.edu}

\maketitle

\begin{abstract}
	Trusted identification is critical to secure IoT devices. However, the limited memory and computation power of low-end IoT devices prevent the direct usage of conventional identification systems.
	RF fingerprinting is a promising technique to identify low-end IoT devices since it only requires the RF signals that most IoT devices can produce for communication.
	However, most existing RF fingerprinting systems are data-dependent and/or not robust to impacts from wireless channels.
	To address the above problems, we propose to exploit the mathematical expression of the physical-layer process, regarded as a function $\mathbf{\mathcal{F}(\cdot)}$, for device identification.
	$\mathbf{\mathcal{F}(\cdot)}$ is not directly derivable, so we further propose a model to learn it and employ this function model as the device fingerprint in our system, namely $\mathcal{F}$ID.
	Our proposed function model characterizes the unique physical-layer process of a device that is independent of the transmitted data, and hence,
	our system $\mathcal{F}$ID is data-independent and thus resilient against signal replay attacks. Modeling and further separating channel effects from the function model makes $\mathcal{F}$ID channel-robust.
	We evaluate $\mathcal{F}$ID on thousands of random signal packets from $33$ different devices in different environments and scenarios, and the overall identification accuracy is over $99\%$.
\end{abstract}


\section{Introduction}
Every Internet-of-Things (IoT) device shall have its own identity to form a trusted ecosystem. Generally, there are two widely-used identification methods for IoT devices, i.e., cryptography-based and hardware-based methods. Cryptography-based scheme provides a unique key for each user or device as the identity. However, all those schemes require the extra computational resource that low-end IoT devices don't have. Hardware-based systems exploit additional hardware to provide security functionalities including identification. Hardware like Intel SGX and TrustZone is a good developing bed for such hardware-based systems \cite{Costan2016}. However, for massively deployed low-end IoT devices, additional hardware is unaffordable. Even for more expensive devices such as laptops and smartphones that already have a cryptography-based or hardware-based identification system, a low-cost identification system can also support multi-factor identification in case that the original system is compromised.

Radio Frequency (RF) fingerprinting is a promising technique to build low-cost identification systems. RF aims at identifying a device by its RF signals, because RF signals reflect the unique hardware imperfections of their source devices which are introduced in the manufacturing process \cite{Danev2012}.
Since almost all IoT devices can produce RF signals for communication and RF fingerprinting only leverages these signals, no additional computational resource and hardware are required by RF fingerprinting systems to be embedded in IoT devices.



However, most existing RF fingerprinting systems are data-dependent and/or not robust to spatial variations and wireless channel effects, mainly including location-based, transient-based, and preamble-based systems.
For location-based systems, the features they use entirely depend on the device's unique location, and hence these systems are sensitive to any spatial variations.
Transient-based and preamble-based systems are typical data-dependent systems since the features they use are extracted from fixed segments of the RF signals, i.e., transition signal or preamble signal. Using a fixed signal segment for identification makes this kind of systems vulnerable to signal replay attacks. An existing partially data-independent and channel-robust RF fingerprinting system is the modulation error-based system \cite{Brik2008}. However, this system completely relies on a 5-feature space for classification. Therefore, the number and types of devices that can be classified by this system are constrained by this low-dimensional feature space.

To address the above problems, we propose a \textbf{F}unction modeling-based
data-Independent and channel-Robust physical-layer \textbf{ID}enti-fication system, namely $\mathcal{F}$ID. We propose to exploit a mathematical function $\mathbf{\mathcal{F}(\cdot)}$ that takes the transmitted data as input and the transmitted RF signal as the output in $\mathcal{F}$ID. $\mathbf{\mathcal{F}(\cdot)}$ is the mathematical expression of the physical-layer process from modulation to power amplification, and hence it can represent all the uniqueness of the hardware and the signal processing procedures within a RF transmitter.
However, $\mathbf{\mathcal{F}(\cdot)}$ is not directly derivable. Hence, we propose an accurate and efficient model, which utilizes several insights about those physical-layer procedures and a widely-used function-learning method (i.e., Kernel Regression) to model $\mathbf{\mathcal{F}(\cdot)}$ for each authenticated device. \textit{\textbf{This function model is employed as the device fingerprint in $\mathcal{F}$ID, i.e., we match the received RF signal and the signal computed/predicted by the function model to identify a device.}}
Since our proposed function and function model characterize the inherent properties of the physical-layer process that remain unchanged regardless of the transmitted data, $\mathcal{F}$ID built on our function model is also data-independent and thus can be resilient against signal replay attacks. Also, the spatial variations and multipath channels can be modeled in our scheme. The impacts of these environmental factors are approximately separable from our function model. Using the remaining part of our proposed function model for identification makes $\mathcal{F}$ID robust to the environmental impacts. Additionally, since $\mathbf{\mathcal{F}(\cdot)}$ can represent all the uniqueness of the hardware and the signal processing procedures within a device's physical-layer process, all the data-independent features are derivable from $\mathbf{\mathcal{F}(\cdot)}$. Therefore, $\mathcal{F}$ID is not constrained by low-dimensional feature spaces, which implies $\mathcal{F}$ID has the potential to identify the devices that can not be classified by the existing feature-based systems.

\textit{\textbf{Contribution.}} Our contributions can be summarized as follows:
\begin{enumerate}
	\item We summarize the limitations of the existing RF fingerprinting systems and propose to exploit the mathematical expression of the physical-layer process (i.e., $\mathbf{\mathcal{F}(\cdot)}$) in our RF fingerprinting system to solve those problems.
	
	\item We propose an accurate and efficient function model to learn $\mathbf{\mathcal{F}(\cdot)}$ since $\mathbf{\mathcal{F}(\cdot)}$ is not directly derivable, and we design a data-independent channel-robust RF fingerprinting system based on our function model, namely $\mathcal{F}$ID.
	
	\item We implement $\mathcal{F}$ID and provide an extensive evaluation to verify the data independency, the outstanding performance, and the robustness of $\mathcal{F}$ID.
\end{enumerate}

The remainder of the paper expands on the above contributions. We begin with brief introduction of the existing RF fingerprint schemes and detailed analysis of their limitations for further explanation of our objectives, followed by the establishment of our proposed function model and the associated $\mathcal{F}$ID system, and evaluation of $\mathcal{F}$ID in different environments and scenarios.

\label{sec:exist_pro}
\begin{figure}[!t]
	\centering
	\includegraphics[width=2in]{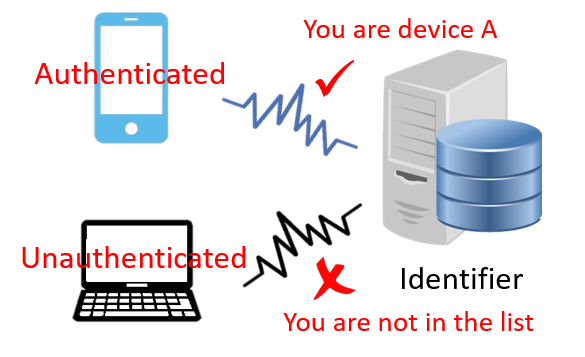}%
	
	\caption{RF fingerprinting (system model)}
	\label{fig:rf_system}
\end{figure}
\section{Problem Statement and Objectives}
\label{sec:problem}
In this section, we first briefly introduce the RF fingerprinting system model. The existing RF fingerprinting schemes and their limitations are summarized in section \ref{sec:relatedWork} and \ref{sec:exist_pro}. To tackle those limitations, we develop a function modeling method and design $\mathcal{F}$ID. Our research objectives are presented in section \ref{sec:motivation}.
\subsection{RF Fingerprinting}
\label{sec:application}
RF fingerprinting is a technique to identify wireless devices by their transmitted RF signals. As illustrated in Fig. \ref{fig:rf_system}, an RF fingerprinting system consists of an identifier and massively deployed wireless devices, which follow certain communication protocols to generate RF signals for communication. The identifier, regarded as a central server in Fig. \ref{fig:rf_system}, is responsible for leveraging the received RF signals and certain algorithms and principles to identify the transmitters of the RF signals.

\subsection{Existing RF Fingerprinting Scheme and System}
\label{sec:relatedWork}

\paragraph{Location-based RF Fingerprinting}
Location-based RF fingerprinting systems are built on features like Radio Signal Strength (RSS) \cite{Zegeye2016}, Channel State Information (CSI) \cite{Tugnait2010}, and Channel Frequency Response (CFR) \cite{Xiao2007} that contain the location information of the target devices.
Therefore, these systems aim to take advantage of the devices' unique locations for device identification.

\paragraph{Transient-based and Preamble-based RF Fingerprinting}
Transient-based and preamble-based RF fingerprinting systems are built on features extracted from the transition signals and preamble signals \cite{Hall2006, Toonstra1996, Shaw1997, Tekbas2004, Kennedy2008, Wang2016, Kennedy2010, Chen2017}. These systems attempt to leverage the uniqueness of a certain fixed segment in all the RF signal packets transmitted by the authenticated devices for device identification.


\paragraph{Modulation Error-based RF Fingerprinting}
A partially data-independent and channel-robust RF fingerprinting system is the modulation error-based system that assigns statistics of the modulation errors as device fingerprint \cite{Brik2008, Zhou2018}. The main five statistics of modulation errors proposed by \cite{Brik2008} include SYNC correlation, carrier frequency offset, averaged magnitude error, averaged phase error and I/Q original offset. Among those five features, carrier frequency offset, SYNC correlation and I/Q original offset are the three most discriminative features \cite{Danev2010}, and SYNC correlation is undoubtedly a data-dependent feature. Therefore, if random RF signals are used for device classification in this system, the carrier frequency offset and I/Q original offset will determine the number and the types of devices that can be classified.


\paragraph{RF Power Amplifier Modeling-based Identification}
In the wireless communication system, the RF power amplifier is a critical hardware component that has been studied for a long time. In previous works, Volterra series and Recurrent Neural Network (RNN) have been successfully used to model the behavior of RF power amplifiers \cite{Zhu2004, Brien2006}. Adam et al. \cite{Adam2011} exploit the Volterra series model to identify different power amplifiers, and they show that the commercial power amplifier chips can be easily identified by very short output sequences. However, to our best knowledge, we are the first to model the whole wireless device rather than a hardware component for device identification. To model such a combination of multiple hardware components, we propose a function model totally different from the Volterra series model.

\paragraph{Deep Learning-based RF Fingerprinting}
Recently, \cite{Kevin2018} and \cite{Rajshekhar2018} exploit convolutional neural network (CNN) and Recurrent Neural Network (RNN) to classify wireless signals for IoT device identification. 
\cite{Kevin2018} demonstrated $92.29\%$ identification accuracy on seven ZigBee devices, and \cite{Rajshekhar2018} achieved over $90\%$ overall accuracy on LoRa low-power wireless chipsets. They are the first trials to apply Deep Learning to device identification (classification), and future work might be needed to reduce their computational cost for further application in reality. Compared with those Deep Learning-based approaches, our model teases apart linear and non-linear effects in the wireless signals rather than a blind use of machine learning. Therefore, our model is more explainable and efficient.

\subsection{Limitations of Existing Work}
\label{sec:exist_pro}
\begin{figure}[!t]
	\centering
	\includegraphics[width=2.8in]{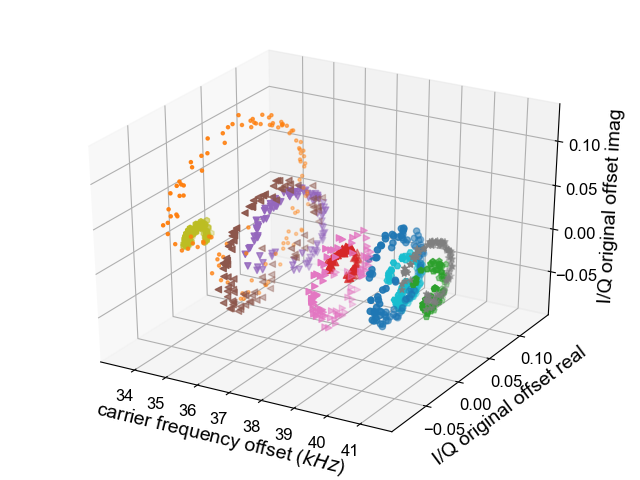}%
	
	\caption{Modulation error feature-space}
	\label{fig:modulationError}
\end{figure}
\paragraph{Data Dependency}
Transient-based and preamble-based RF fingerprinting systems are typical data-dependent systems, because each time these systems only use the same signal segment (i.e., the transient signal and/or the preamble signal) for identification. A significant weakness of data-dependent RF fingerprinting systems is that they are vulnerable to signal replay attacks in which the attackers simply record the RF signals from an authenticated device and replay the same signal segment to the server to impersonate the authenticated devices.

\paragraph{Robustness}
Most existing RF fingerprinting systems are not robust to spatial variations and/or channel effects. These systems mainly include location-based systems, transient-based systems, and preamble-based systems. For location-based systems, what these systems really identify is the device's unique location, hence these systems are sensitive to any spatial variations. For transient-based and preamble-based systems, the transient-based and preamble-based features are always derived by spectral transformations, such as Fast Fourier Transform or Discrete Wavelet Transform. These kinds of features are proved to be sensitive to distance and orientation variations \cite{Danev2009, Danev2010}.

\paragraph{Constrained Feature Space}
Most existing RF fingerprinting systems completely rely on certain features and their performance is constrained by the associated low-dimensional feature space. For instance, the performance of the modulation error-based system is mainly determined and thus constrained by the aforementioned two features, i.e., carrier frequency offset and I/Q original offset. In order to justify our statement, we randomly select 10 telosb sensors and plot those two features computed by the random signal packets collected from these 10 sensors in Fig. \ref{fig:modulationError}. Here we use 10 colors to represent those 10 sensors. Each point represents those two features computed by a random signal packet. As shown in Fig. \ref{fig:modulationError}, I/Q original offset is also a data-dependent feature since it shows a significant difference between different random signal packets from one device. Therefore, in this case, the carrier frequency offset is the only determinant, and it is unable for the modulation error-based system to distinguish between several sensors only by their carrier frequency offsets. Specifically, two pairs of sensors are indistinguishable in this five-feature space
(i.e., actually only a single-feature space).

\subsection{Research Objectives}
\label{sec:motivation}
Our core objective is to design an RF fingerprinting system that can get rid of the above limitations. To this end, we aim to design a scheme/system that has the following characteristics: First, our system should be able to identify the IoT devices by the random RF signal packets collected from those devices, and hence it can be completely immune to signal replay attacks with the help of a challenge and response protocol. Second, our system should also be robust to spatial variations and multipath channels so that it can be applied in reality. Third, our system should not be constrained by a certain low-dimensional feature space, or in another word, our system should have the potential to identify the IoT devices that are indistinguishable in any low-dimensional feature space.

\section{$\mathcal{F}$ID Function Modeling for Fingerprinting}
\label{sec:funcModel}

\begin{figure}[!t]
	\centering
	\includegraphics[width=2.8in]{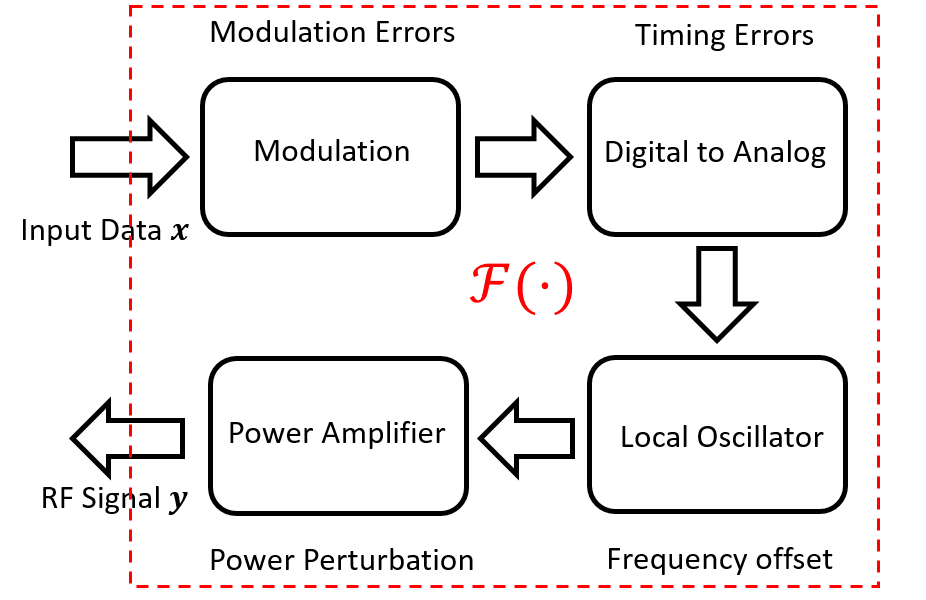}%
	
	\caption{The process of how the input data is transformed into the output RF signal}
	\label{fig:process}
\end{figure}

In $\mathcal{F}$ID, we propose to exploit a mathematical function $\mathbf{\mathcal{F}(\cdot)}$ that takes the transmitted data $\mathbf{x}$ as input and outputs the RF signal $\mathbf{y}$, i.e., $\mathbf{y} = \mathbf{\mathcal{F}(x)}$. \textit{\textbf{As shown in Fig. \ref{fig:process}, $\mathbf{\mathcal{F}(\cdot)}$ is the mathematical expression of the physical-layer process from modulation to power amplification. Hence, it can represent all the impacts of hardware imperfections, including modulation errors, timing errors, frequency offset and power perturbation, within the physical-layer process of a wireless device.
		Moreover, $\mathbf{\mathcal{F}(\cdot)}$ is apparently independent of the transmitted data and the external wireless channels.
		Therefore, using the expression/model of $\mathbf{\mathcal{F}(\cdot)}$ to identify a wireless device can realize all the aforementioned objectives.}}
However, to our best knowledge, $\mathbf{\mathcal{F}(\cdot)}$ is not directly derivable, and there is also not a well-developed model to learn $\mathbf{\mathcal{F}(\cdot)}$. Therefore, in section \ref{sec:nonfunc}, we first propose an accurate and efficient model to learn $\mathbf{\mathcal{F}(\cdot)}$ and employ this function model as the device fingerprint in $\mathcal{F}$ID. In section \ref{sec:modelingChannel}, we further model the spatial variations and multipath and mobile channels to mitigate the impacts of those environmental factors.

\subsection{Function Modeling of Hardware Imperfections}
\label{sec:theory}

\label{sec:nonfunc}

\textit{\textbf{In our function model, instead of directly modeling $\mathbf{\mathcal{F}(\cdot)}$, we choose to model a simpler intermediate function $f(\cdot)$. The input of this intermediate function is transformed from the transmitted data into the ideal signal}}, which is defined as the baseband signal generated by the input data $x$ and an imaginary transmitter without hardware imperfection. The output is still the RF signal $\mathbf{y}$ transmitted by the wireless device.
\textit{\textbf{Modeling $f(\cdot)$ is equivalent to modeling $\mathbf{\mathcal{F}(\cdot)}$, since the transformation from the transmitted data to the ideal signal is already defined by the communication protocol.}}, i.e., $\mathbf{\mathcal{F}(\cdot)} = f(T(\cdot))$, where $T(\cdot)$ is known in advance. Moreover, it is more convenient to use our insights about the hardware imperfections if modeling $f(\cdot)$. 

\paragraph{Definition and Notation} We regard the ideal signal and output RF signal as $x(t)$ and $y(t)$ respectively. \textit{\textbf{Hence, we have $y(t) = f(x(t))$. This $f(\cdot)$ is the continuous form of the immediate function.}} We assume that the signals are sampled at $kT_s+\tau (k = 0, 1, 2 ...)$, where $T_s$ is the sampling interval and $\tau$ is the sampling phase. Their $n^{th}$ samples are denoted by $x[n]$ and $y[n]$. \textit{\textbf{The relationship between $y[n]$ and $x(t)$ or $x[n]$s is defined as the discrete form of the immediate function.}}

\paragraph{$f(\cdot)$'s discrete form}Our first insight is that $y[n]$ is affected not only by $x[n]$, but also by the signal on both sides of $x[n]$, i.e., $y[n]$ is the function of a segment of the ideal signal, $\{x(t)|(n-k)T_s < t < (n+m)T_s\}$ and the sampling phase, $\tau$. Here we regard $\{x(t)|(n-k)T_s < t < (n+m)T_s\}$ as $x_n(t)$. Then, the discrete form of $f(\cdot)$ can be expressed as
\begin{equation}
y[n] = f(x_n(t), \tau).
\end{equation}

As a sample in RF signal, y[n] can be expressed as $Ae^{i\theta}$, thus we can further express the discrete form of $f(\cdot)$ as
\begin{equation}\label{eq:interfunc1}
f(x_n(t), \tau) = A^t(x_n(t), \tau)e^{iw_c^tt + i\theta^t(x_n(t), \tau)}.
\end{equation}
Here $A^t(x_n(t), \tau)$ is the magnitude of $y[n]$ and $w_c^tt + \theta(x(t), \tau)$ is the phase of $y[n]$. $w_c^t$ is the carrier frequency of the transmitter. Our second insight is that a carrier frequency offset exists due to the imperfections of the local oscillator. Hence, $f(x_n(t), \tau)$ can be rewritten as
\begin{equation}\label{eq:interfunc2}
f(x_n(t), \tau) = A^t(x_n(t), \tau)e^{iw_ct + \Delta w^tt + i\theta^t(x_n(t), \tau)},
\end{equation}
where $\Delta w^t$ is the offset.

\paragraph{Decomposition of $f(\cdot)$}
In order to model different portions of the function $f(\cdot)$ accurately and efficiently, we decompose the amplitude and the phase into linear parts and nonlinear parts:
\begin{equation}\label{eq:interfunc3}
\begin{aligned}
f(x_n(t), \tau) &=  A_0^t(1 + pow^t(x_n(t), \tau))      \cdot                                 \\
& e^{iw_c(nT_s+\tau) + i\Delta w^t(nT_s+\tau) + i\theta(nT_s+\tau) + i\Theta^t(x_n(t), \tau)}   ,\\
\end{aligned}
\end{equation}
where $pow^t(x_n(t), \tau)$ is the nonlinear part of the amplitude and $\Theta^t(x_n(t), \tau)$ is the nonlinear part of the phase. $\theta(nT_s+\tau)$ is added in the modulation stage.

\paragraph{$f(\cdot)$ in reality} In reality, it is hard to know the exact output RF signals of a device. Strictly speaking, what we can collect are only the RF signals received by an RF receiver. To address this problem, we can refine $\mathbf{f}(\cdot)$ as a function whose input is the ideal signal and output is the received signal. Similar to Eq. \ref{eq:interfunc3}, $\mathbf{f}(\cdot)$ can be expressed as
\begin{equation}\label{eq:interfunc4}
\begin{aligned}
f(x_n(t), \tau) &=  A_0^r(1 + pow^r(x_n(t), \tau))      \cdot                                 \\
& e^{i\theta_0 + i(\Delta w^t - \Delta w^r)(nT_s+\tau) + i\theta(nT_s+\tau) + i\Theta^r(x_n(t), \tau)}.\\
\end{aligned}
\end{equation}
In this equation, the extra term $e^{i\theta_0}$ is the channel coefficient.
$\Delta w^r$ is the carrier frequency offset of the receiver. The nonlinear terms $pow^r(x_n(t), \tau)$ and $\Theta^r(x_n(t), \tau)$ are are the nonlinear parts of the received signal, determined by both the transmitter and the receiver.

\paragraph{Kernel regression}Based on Eq. \ref{eq:interfunc4}, our problem is reduced from modeling a "black box" to modeling two nonlinear terms, i.e., $pow^r(x_n(t), \tau)$ and $\Theta^r(x_n(t), \tau)$. This is because all the other linear terms can be directly computed, which is clarified in section \ref{sec:demodulation}. We propose to use Kernel Regression to learn those two nonlinear terms.
In order to implement Kernel Regression, some representative digital samples in the signal segment $x_n(t)$ are used to replace $x_n(t)$ as the input vector. These samples are $[x'[n-k], x'[n-k+1], ..., x'[n+m-1], x'[n+m]]^T$ where $x'[n]$ is the digital sample sampled at $nT_s$ in the ideal signal. \textbf{\textit{According to Nyquist Theorem, once the sampling period $T_s$ is smaller than half of the symbol period, these digital samples are equivalent to $x_n(t)$. Hence, the input vector can expressed as
		\begin{equation}\label{eq:trainingvector}
		[x'[n-k], x'[n-k+1], ..., x'[n+m-1], x'[n+m], \tau]^T,
		\end{equation}
		and the target values are those two nonlinear terms.}}
Section \ref{sec:experiment} shows that if our function model is applied, the modeling accuracy can be very high and the computational cost is also acceptable for a commodity server.



\subsection{Function Modeling of Environmental Factors}
\label{sec:modelingChannel}

\paragraph{Modeling Spatial Variations}
Here we model two main spatial variations, i.e., communication distance and orientation variations.
Varying communication distance changes the amplitude of the received signal, then $f(x_n(t), \tau)$  will become
\begin{equation}\label{eq:distance}
\begin{aligned}
f(x_n(t), \tau) = & A^r(d)(1 + pow^r(x_n(t), \tau))      \cdot                                 \\
& e^{i\theta_0 + i(\Delta w^t - \Delta w^r)(nT_s+\tau) + i\theta(nT_s+\tau) + i\Theta^r(x_n(t), \tau)}   ,\\
\end{aligned}
\end{equation}
where $A^r$ is not a constant but a function of communication distance $d$. In this case, our function model is still workable for device identification, since the other linear terms and two normalized nonlinear terms will not be significantly affected by $d$, and those terms can be used for device identification.

Another spatial variation considered here is the communication orientation variation. Communication orientation refers the polarization mismatching angle between the transmitter's antenna and the receiver's antenna. The impact of the polarization mismatch is equivalent to multiplying the received RF signal by a projection factor, $cos\alpha$. $\alpha$ is the mismatching angle between the transmitter's and the receiver's antennas. The impact of orientation variation is similar to the impact of varying the communication distance, and hence it can be addressed in the same way.

\paragraph{Modeling Multipath Channel}
\label{sec:multipathChannel}
Multipath channel is always caused by signal reflection and refraction. If multiple paths occur in the wireless communication process, then the received signal can be seen as the summation of RF signals coming from these paths. A conventional method to model the multipath channel is channel estimation, where the received signal $Z[n]$ is expressed as
\begin{equation}\label{eq:multipath}
Z[n] \approx \sum_{i=1}^{N}h[i] \cdot f(x_{n-i}(t), \tau).
\end{equation}
Here $h[i]$ can be approximated by linear regression. After the channel taps $h[i]$ are computed, we can deconvolve $f(x_n(t), \tau)$ out of $Z[n]$ for device identification.

\section{$\mathcal{F}$ID System Design}
\label{sec:system}

\begin{figure}[!t]
	\centering
	\includegraphics[width=3.2in]{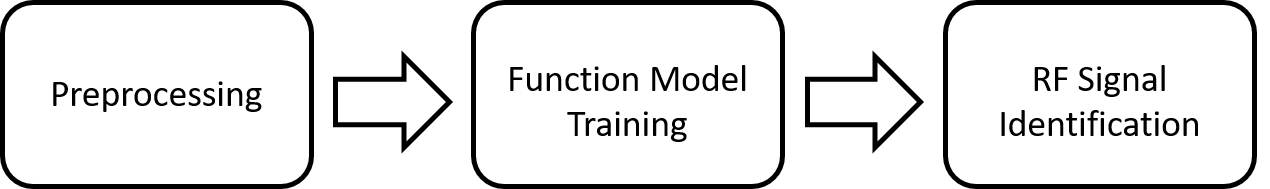}%
	\caption{System Modules}
	\label{fig:system}
\end{figure}

$\mathcal{F}$ID is designed based on the function model derived in section \ref{sec:nonfunc}. As shown in Fig. \ref{fig:system}, our system consists of 3 modules. The first module (i.e., Preprocessing Module) is used to extract the linear terms and parameters from the received signal. The second module (i.e., Function Model Training Module) is used to train the function model. The last module (i.e., RF Signal Identification Module) utilizes the results from the first and second module to identify the received RF signal.

\subsection{Preprocessing Module}
\label{sec:demodulation}
\begin{figure}[!t]
	\centering
	\includegraphics[width=3.2in]{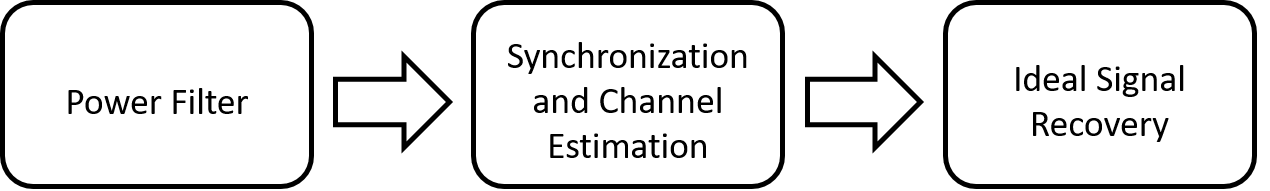}%
	
	\caption{Preprocessing Submodules}
	\label{fig:demodulation}
\end{figure}

\begin{figure*}[!ht]
	\centering
	\includegraphics[width=5in]{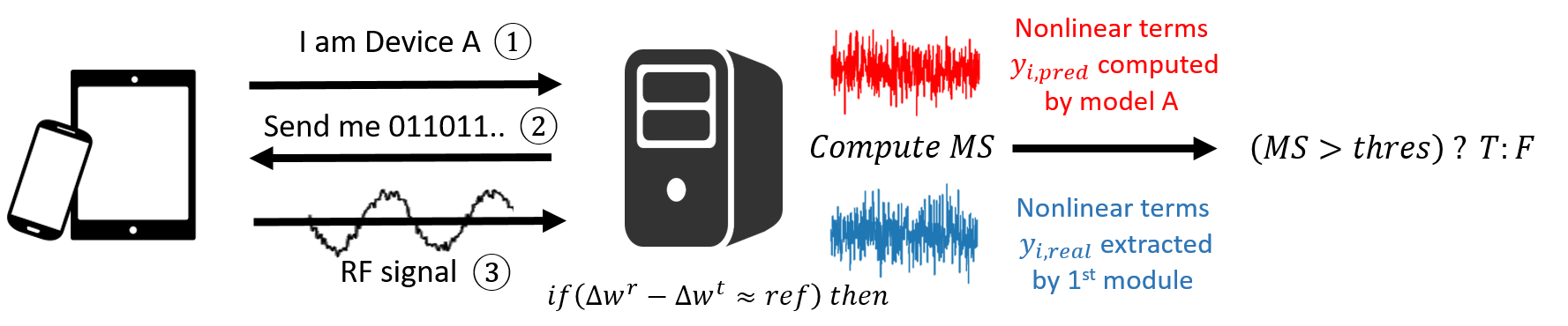}%
	
	\caption{Identification procedures}
	\label{fig:identificationmodule}
\end{figure*}
For the first module, the input is the received signal and the outputs include the digital samples of the ideal signal $x'[n]$, sampling phase $\tau$, channel coefficient $e^{i\theta}$ or channel taps $h[n]$, and carrier frequency offset. This module consists of 3 submodules as shown in Fig. \ref{fig:demodulation}. Power filter is used to find the start and the end of each signal packet. This submodule compares the absolute value of every received sample with a power threshold to localize each signal packet.

The second submodule is used to synchronize the received packet to compute the \textit{\textbf{carrier frequency offset}}. We propose an algorithm, where a modified phase locked loop is used to realize synchronization. \textit{\textbf{We regard the deconvolved received samples as $\mathbf{z[n]}$ }} and compute the phase difference between two adjacent received digital samples $\Delta\theta_z[n]$ by
\begin{equation}\label{eq:phasediff}
\Delta\theta_z[n] = \angle(z^*[n]z[n+1]).
\end{equation}
Based on the eq. \ref{eq:interfunc4}, $\Delta\theta_z[n]$ can also be expressed as
\begin{equation}\label{eq:phasediff2}
\begin{aligned}
\Delta\theta_z[n] = &(\Delta w^t - \Delta w^r)T_s + \theta((n+1)T_s+\tau) - \theta(nT_s+\tau) \\
& + \Theta^r(x_{n+1}(t), \tau) - \Theta^r(x_n(t), \tau),
\end{aligned}
\end{equation}
where $\theta((n+1)T_s+\tau) - \theta(nT_s+\tau)$ is the phase change between two adjacent ideal digital samples. Since
\begin{equation}\label{eq:nonphasediff}
E(\Theta^r(x_{n+1}(t), \tau) - \Theta^r(x_n(t), \tau)) = 0,
\end{equation}
the carrier frequency offset can be computed by
\begin{equation}\label{eq:freqoffset}
\begin{aligned}
\Delta w^t - \Delta w^r = [E(\angle(z^*[n]z[n+1]) - \angle(x^*[n]x[n+1]))]/T_s,
\end{aligned}
\end{equation}
where $\angle(x^*[n]x[n+1]) = \angle(x'^*[n]x'[n+1])$. The $x'[n]$s in the preamble are known in advance and thus used to calculate the expectation, i.e., $E(\angle(z^*[n]z[n+1]) - \angle(x^*[n]x[n+1])$. 

Finally, the ideal digital samples following the preamble are computed. We first compute the ideal phase difference between the following ideal samples, i.e., $\theta((n+1)T_s) - \theta(nT_s)$. Here $\theta((n+1)T_s) - \theta(nT_s)$ can be determined by its approximation, i.e., $\angle(z^*[n]z[n+1]) - (\Delta w^t - \Delta w^r)T_s$:
In the commonly-used communication protocols, there are $2 \sim 5$ possible values for $\theta((n+1)T_s) - \theta(nT_s)$ if the $T_s$ is fixed. For instance, for the protocols using BPSK (e.g., $IEEE 802.11$), there are $3$ possible values, i.e., $0$ and $\pm \pi$. For the protocols using QPSK, there $5$ possible values, i.e., $0$, $\pm \frac{\pi}{2}$ and $\pm \pi$. For the protocols using OQPSK (e.g., $IEEE 802.15.4)$, there are $2$ possible values and they are opposite numbers, which depend on $T_s$. Hence, given a protocol and $T_s$, $\theta((n+1)T_s) - \theta(nT_s)$ is determined as the possible value closest to $\angle(z^*[n]z[n+1]) - (\Delta w^t - \Delta w^r)T_s$. Besides, in the protocols using OQPSK, two consecutive samples might be sampled at both sides of a transition point (i.e., the intersection point of two adjacent symbol periods) and the distances between them and the transition point can be very similar. Therefore, the phase difference between those two consecutive samples is close to $0$ (i.e., the middle of the $2$ possible values), and then a correct decision can not be guaranteed. To tackle this problem, we could double the sampling frequency and make the decision by the phase difference between the digital samples at $nT_s+\tau$ and $(n+2)T_s+\tau$ using the same method. After obtaining $\theta((n+1)T_s) - \theta(nT_s)$, the following ideal samples can be easily computed by $e^{i(\sum_{i=p+1}^{i=n}\theta(iT_s) - \theta((i-1)T_s) + \theta(p))}$, where $\theta(p)$ is the phase of the last digital sample in the preamble.

After these operations, we obtain all the linear parts in the received signal. Some of these linear parts serve as the input for the next two modules. Some unique data-independent linear parameters (e.g., carrier frequency offset) are used for device identification.
To check the correctness of this module, we demodulate the transmitted data using this module, and the bit error is less than $10^{-5}$.

\subsection{Function Model Training Module}

Since all the linear parts can be directly computed by the first module, here only the nonlinear terms need to be learnt to establish the function model, i.e., Eq. \ref{eq:interfunc4}. A widely-used function-learning method, i.e., Kernel Regression (KR), is incorporated to learn those two nonlinear terms in this stage. To train the KR models, we use the outputs of the first module, i.e., the ideal digital samples $x'[n]$ and sampling phase $\tau$, to construct training vectors (i.e., input vectors). The target values are those nonlinear terms. Linear Kernel, Polynomial Kernel, and Radial Basis Function (RBF) kernel are tested, and RBF Kernel model provides the best performance.
Aside from training the KR models, this module also serves as a database to seal the function model, including the KR models and those linear parameters.

\subsection{RF Signal Identification Module}

\label{sec:modeltest}
This module leverages the first two modules to identify the received RF signal. Specifically, the identification principles and procedures are incorporated in this module.
\textit{\textbf{In order to identify the received RF signal, this module matches the received RF signal with the signal computed/predicted by the function model. In practice, we only match the carrier frequency offset and those two nonlinear terms. This is because all the other linear parts are highly related to the transmitted data and/or the environment as illustrated in section \ref{sec:funcModel}.}}

We define a metric named matching score ($MS$) as
\begin{equation}\label{eq:score}
MS = 1-\sum_{i=1}^{N}(y_{i,pred}-y_{i,true})^2/\sum_{i=1}^{N}(y_{i,true}-\overline{y_{true}})^2
\end{equation}
to evaluate the similarity between the real nonlinear terms and the predicted (computed) nonlinear terms. If $y_{i,pred}$ equals $y_{i,true}$, then the $MS$ is 1. If they are totally different, then the $MS$ can be negative. $MS$ can also interpreted as a metric to evaluate the accuracy of the predicted nonlinear terms.

The detailed procedures for device identification are shown in Fig. \ref{fig:identificationmodule}. A challenge and response protocol is applied here: Every time a device wants to be identified, it first sends a request containing its identity (e.g., Device A) to the identification server. Next, the server generates a random data sequence and send it to the device. Then the device needs to modulate the data and send the data back to the server by RF signals. After the server receives the RF signal, the first module of $\mathcal{F}$ID is used to extract linear parts and nonlinear terms and try to match the carrier frequency offset (i.e., $\Delta w^t - \Delta w^r$) with the reference. If $\Delta w^t - \Delta w^r$ is matched, then $\mathcal{F}$ID computes the nonlinear terms by the pre-trained function model (e.g., model A) and the $MS$ between the computed nonlinear terms and the extracted nonlinear terms. Finally, we compare the $MS$ with a predefined threshold (e.g., 0.9) for identification. If the $MS$ is higher than this predefined threshold, then the received signal is identified as coming from the genuine device (e.g., Device A). Otherwise, the received signal is considered as coming from an unauthenticated device. This predefined threshold can be adjusted based on the required accuracy, the communication environment, and the similarity between devices.


\section{Experiments and Analysis}
\label{sec:experiment}

\begin{figure}[!t]
	\centering
	\includegraphics[width=2.0in, height=1.6in]{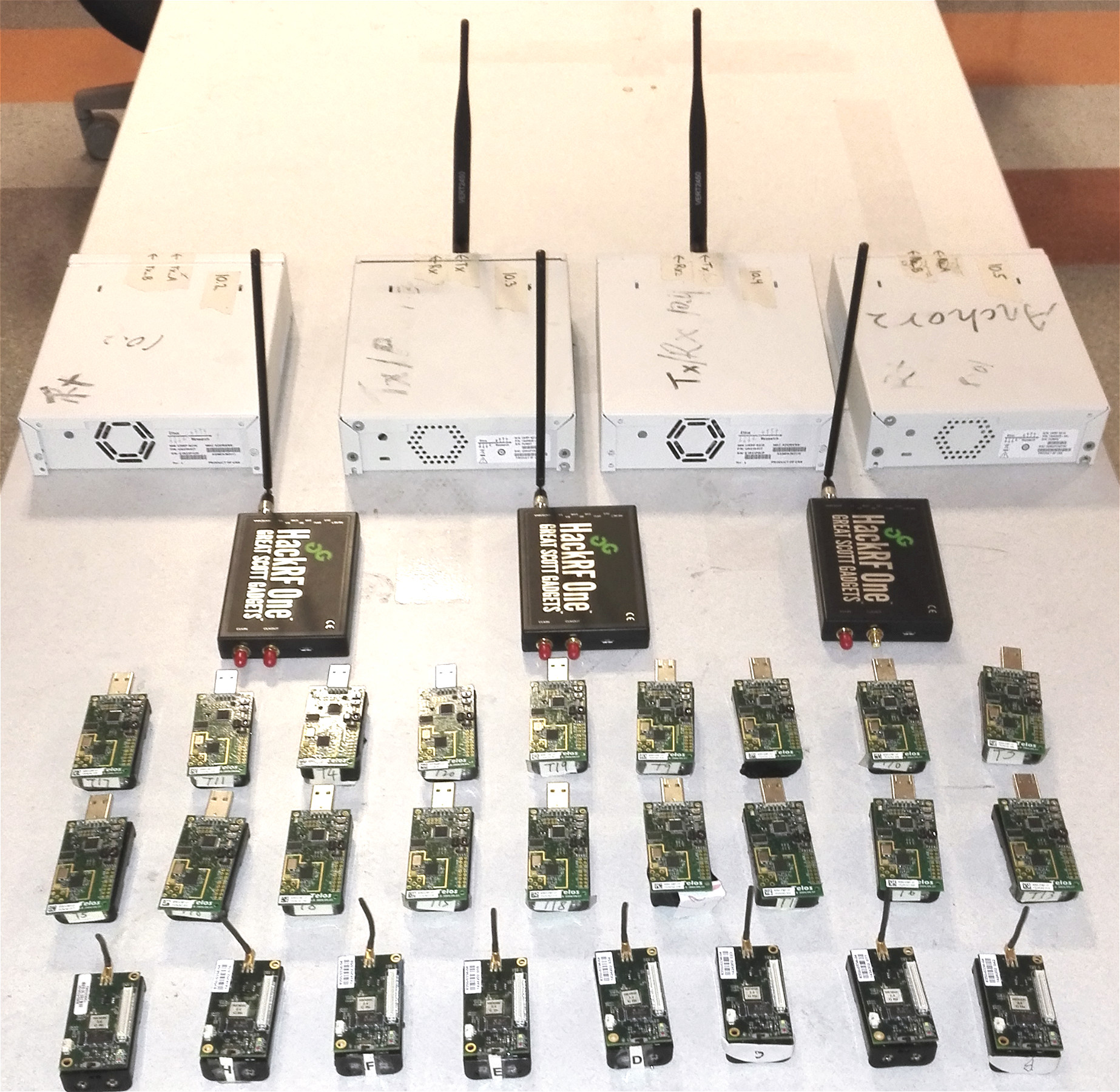}%
	
	\caption{Tested devices}
	\label{fig:experimentDevice}
\end{figure}

We implemented $\mathcal{F}$ID on an Ubuntu16.04.2 PC with an Intel Core i5-2400 CPU @ 3.10GHz processor, and this PC is connected to an Ettus USRP transceiver to form an identification server. $\mathcal{F}$ID is tested on multiple types of devices, including high-end devices like Software Defined Radio (SDR) transmitters (i.e., Ettus USRP and HACKRF) and low-end devices like micaz and telosb sensors, as shown in Fig. \ref{fig:experimentDevice}. Specifically, for the SDR transmitters, the modulation scheme is Quadrature Phase-Shift Keying (QPSK), and the RF center frequency and symbol rate are configured as $2.4Gz$ and $2M/s$ respectively. For low-end Zigbee sensors, the communication protocol is $IEEE802.15.4$, where the symbol rate is $1M/s$ and the RF center frequency is configured as $2.48Gz$.
Here the sampling rate of the receiver (i.e., Ettus USRP transceiver) connected to the PC is configured as $4M/s$. To verify the data independency of $\mathcal{F}$ID, transmitted data is generated by the software-based random number generators developed in the Gnuradio (for SDR) and the TinyOS (for Zigbee devices).

The results in section \ref{sec:modelEvaluation} demonstrate that $\mathcal{F}$ID can model those two nonlinear terms with high accuracy and efficiency. And since all the other linear parts can be directly computed by the preprocessing module, we can say that our function model is a high-precision model.

In order to evaluate the identification performance of $\mathcal{F}$ID, $4$ metrics are applied, including Genuine Acceptance Rate (GAR), Genuine Rejection Rate (GRR), False Acceptance Rate (FAR), and False Rejection Rate (FRR). GAR/GRR refer to the rate at which $\mathcal{F}$ID succeed/fail to identify the genuine device using its function model. FAR/FRR refer to the rate at which $\mathcal{F}$ID accepts/rejects other devices using the genuine device's function model.
In the experiments, each time we choose one device as the genuine device and test the signal packets from it and other devices by this genuine device' function model. Then we compute the GAR, GRR, FAR, and FRR for $33$ different devices. We define Balanced Identification Accuracy (BIA) as
\begin{equation}\label{eq:evaluation}
BIA = \frac{GAR + FRR}{GAR + FAR + GRR + FRR}
\end{equation}
to evaluate the overall performance of $\mathcal{F}$ID.
The experiment results show that BIA is $100\%$ in the line-of-sight environments and over $99\%$ in the multipath environments. We also show that $\mathcal{F}$ID is able to identify the sensor nodes that the modulation error-based system can not classify.

\subsection{Function Model Evaluation}
\label{sec:modelEvaluation}
%
%
%


\paragraph{Function Modeling Accuracy}
\begin{figure}[htpb]
	\centering
	\subfloat[$pow^r(x_n(t), \tau)$]{
		\includegraphics[width=0.48\columnwidth]{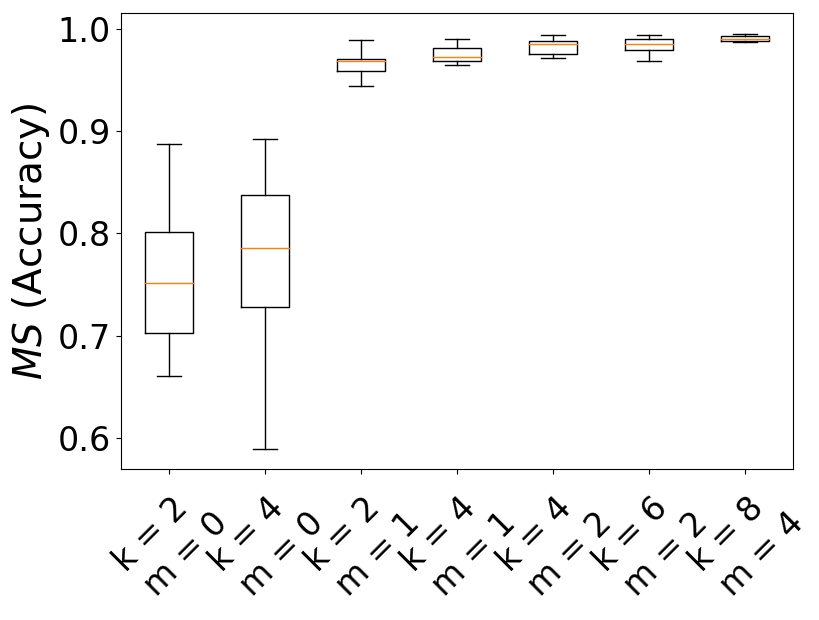}
		\label{fig:powaccuracytelosb}
	}
	\subfloat[ $\Theta^r(x_n(t), \tau)$]{
		\includegraphics[width=0.48\columnwidth]{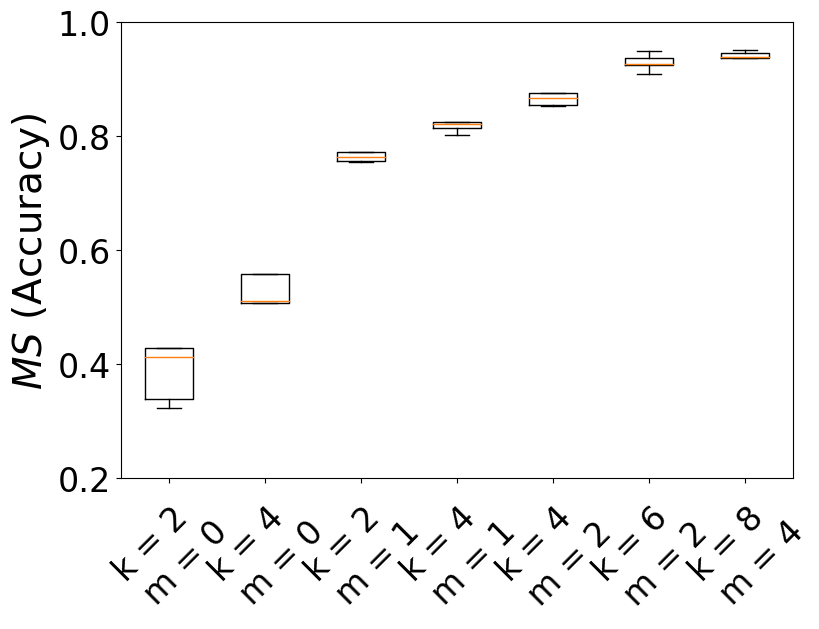}
		\label{fig:phaseaccuracytelosb}
	}
	\caption{Modeling accuracy of those two KR models}
	\label{fig:modeling_accuracy}
\end{figure}

Since all the linear parts in the collected RF signals can be simply computed by the first module of $\mathcal{F}$ID, our task is reduced to verifying the accuracy of those $2$ KR models for $pow^r(x_n(t), \tau)$ and $\Theta^r(x_n(t), \tau)$.
Aside from verifying that our model is a high-precision model, we also want to study the impacts of those two parameters in the input vector (Eq. \ref{eq:trainingvector}), i.e., $k$ and $m$, on the modeling accuracy.
\textit{\textbf{We regard the matching score between the real and predicted power nonlinear term as $MS_{pow}$, and the matching score between the real and predicted phase nonlinear term as $MS_{\Theta}$.}}
Figure \ref{fig:powaccuracytelosb} and \ref{fig:phaseaccuracytelosb} display the modeling accuracy by the statistics of the $MS_{pow}$ and $MS_{\Theta}$ of the testing signals. \textit{\textbf{Here all the testing signals are collected from those $33$ devices and tested only by their own function models.}} 
Figure \ref{fig:powaccuracytelosb} shows that the highest testing $MS_{pow}$ is nearly $0.99$. Considering the existence of the ambient noise, it is hard to improve this result even with a much more complicated model. It is also shown that once $k \geq 2$ and $m \geq 1$, it is enough for our model to capture most information in the power nonlinear term.
Figure \ref{fig:phaseaccuracytelosb} shows that the highest testing $MS_{\Theta}$ is above $0.9$ and once $k \geq 6$ and $m \geq 2$, our model is able to capture most information in the phase nonlinear term. In the following experiments, we set $k = 8$ and $m = 4$ to train the KR models.

\paragraph{Function Modeling Efficiency}
\begin{figure}[!t]
	\centering
	\includegraphics[width=2.4in]{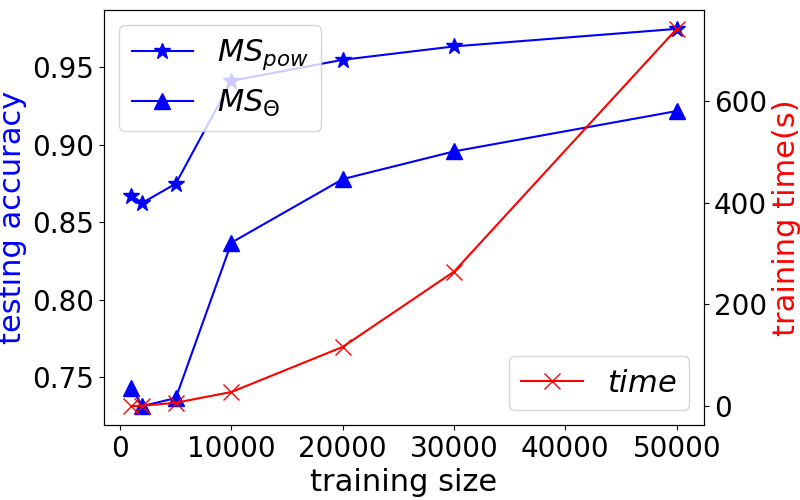}%
	
	\caption{Testing accuracy and training time}
	\label{fig:efficiency}
\end{figure}

\label{sec:modelaccuracy}

During the experiments, we found that the training time of a function model highly depends on the training data size, i.e., number of digital samples. Therefore, we tune the training data size and plot the corresponding testing accuracy and training time in Fig. \ref{fig:efficiency}. When the training data size is larger than $40000$, the modeling accuracy is enough for device identification. The averaged time for training a function model is less 10 minutes when the training data size is 40000. Compared with RNN that takes tens of hours or even more than a day to learn $f(\cdot)$, our function model is undoubtedly more time-efficient.

\subsection{Data Independency Verification}
\label{sec:data_independent_verify}
\begin{figure}[htpb]
	\centering
	\subfloat[preambles]{
		\includegraphics[width=0.48\columnwidth]{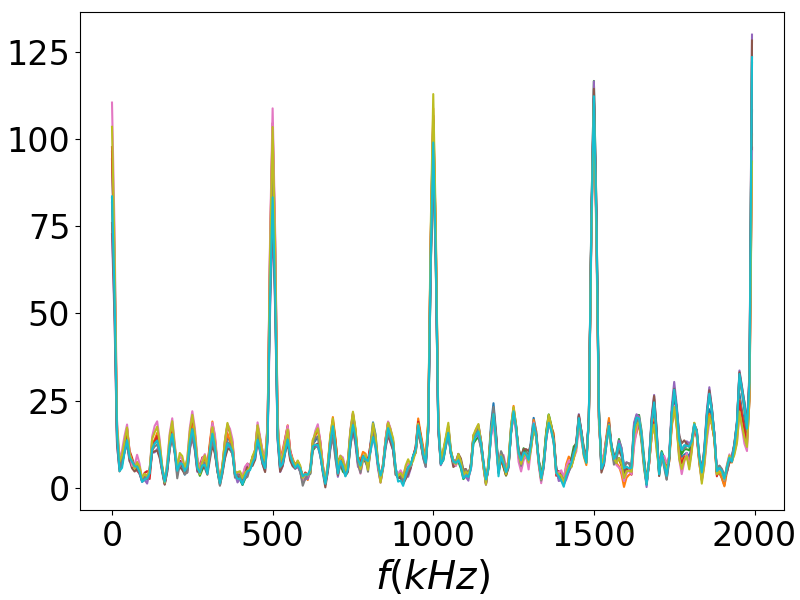}
	}
	\subfloat[random packets]{
		\includegraphics[width=0.48\columnwidth]{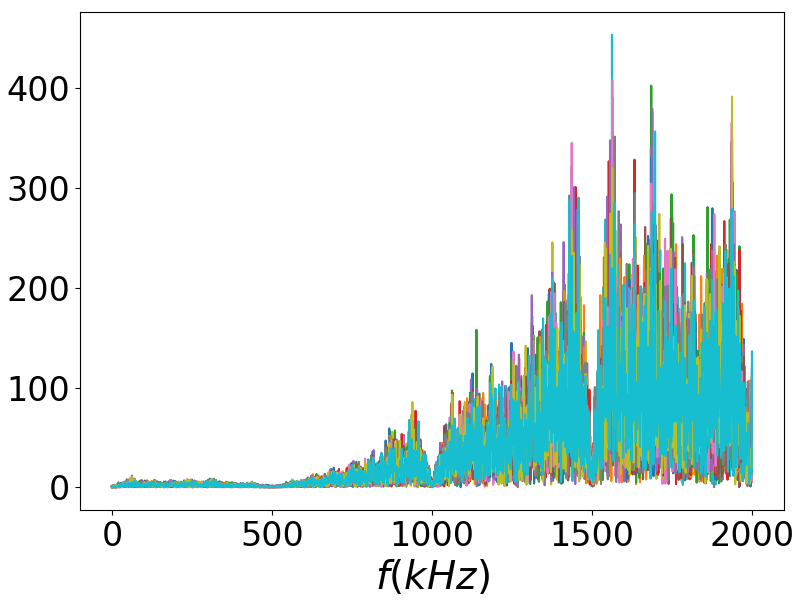}
	}
	\caption{FFT Spectrums of the 10 preambles and 10 random packets collected from one device}
	\label{fig:prea_packet_corr}
\end{figure}

As stated above, all input data is generated by software-based random number generators, and the preamble part is dropped from the received RF signals. Therefore, all the RF signal packets for training, testing, and identification are random signal packets. To prove this, we compute the correlation between the collected signal packets. The average of the correlation coefficients is $0.18 \sim 0.19$, and variance is approximately $0.03$. For comparison, we also compute the correlation between the preamble signals, and the average of the correlation coefficients is $0.98 \sim 0.99$. Besides, we also randomly select 10 signal packets and 10 preambles collected from one device and plot the FFT spectrums of these preamble signals and random signal packets in Fig. \ref{fig:prea_packet_corr}. We found the FFT spectrums of the preamble signals are very similar, but the FFT spectrums of these 10 random signal packets are distinct despite coming from the same device.
These results indicate that the signal packets we utilize for identification are random signal packets. Therefore, the attackers can not replay those packets under the challenge and response protocol introduced in section \ref{sec:modeltest}. In another word, even the adversaries can record the RF signals from authenticated devices with a high-end RF transceiver, but they can not replay those signals for identification as long as the "challenge" changes each time.

\begin{table*}
	\caption{Comparison between $\mathcal{F}$ID and modulation error-based system (in line-of-sight environments) }
	\center
	\begin{tabular}{ |p{2cm}|p{1.3cm}|p{1.3cm}|p{1.3cm}|p{1.3cm}|p{1.3cm}|p{1.3cm}|p{1.3cm}|p{1.3cm} }
		\hline
		\centering
		&\multicolumn{4}{|c|}{$\mathcal{F}$ID}&\multicolumn{4}{|c|}{Modulation error-based System} \\
		\hline
		& \multicolumn{2}{|c|}{overall performance} & \multicolumn{2}{|c|}{two most similar devices}& \multicolumn{2}{|c|}{overall performance} & \multicolumn{2}{|c|}{two most similar devices} \\
		\hline
		GAR, GFR & 0.97, 0.03 & 1.0, 0.0 & 0.94, 0.06 & 1.0, 0.0& \multicolumn{2}{|c|}{0.91, 0.09} & \multicolumn{2}{|c|}{0.58, 0.42} \\
		\hline
		FAR, FFR & 0.0, 1.0 & 0.0, 1.0 & 0.0, 1.0 & 0.0, 1.0& \multicolumn{2}{|c|}{0.12, 0.88} & \multicolumn{2}{|c|}{0.44, 0.56}\\
		\hline
		BIA & 0.99 & 1.0 & 0.97 & 1.0 & \multicolumn{2}{|c|}{0.90} & \multicolumn{2}{|c|}{0.57}\\
		\hline
	\end{tabular}
%
	\label{tab:performance}
\end{table*}
\subsection{Identification Performance Evaluation}
\label{sec:lineofsight}

To show the performance of $\mathcal{F}$ID, we first try to identify the RF signals collected in the line-of-sight environments, where the communication distance and orientation can vary. Here we use the method introduced in section \ref{sec:modeltest} to identify the received RF signals. The thresholds for $MS_{pow}$ and $MS_{\Theta}$ are set as $0.94 \sim 0.95$ and $0.9$ respectively based on the function modeling results.
Since all the experiments are conducted indoor with uncertain ambient noise (e.g., ambient wifi signals), for every $10 \sim 20$ signal packets from the genuine device, there might exist one bad packet. To address this problem, we make the identification decision by testing two adjacent packets, and if one of them results in $MS$s higher than those two thresholds, then the request is accepted.

\begin{figure}[!h]
	\centering
	\begin{minipage}[t]{0.48\columnwidth}
		\includegraphics[width=1.0\columnwidth]{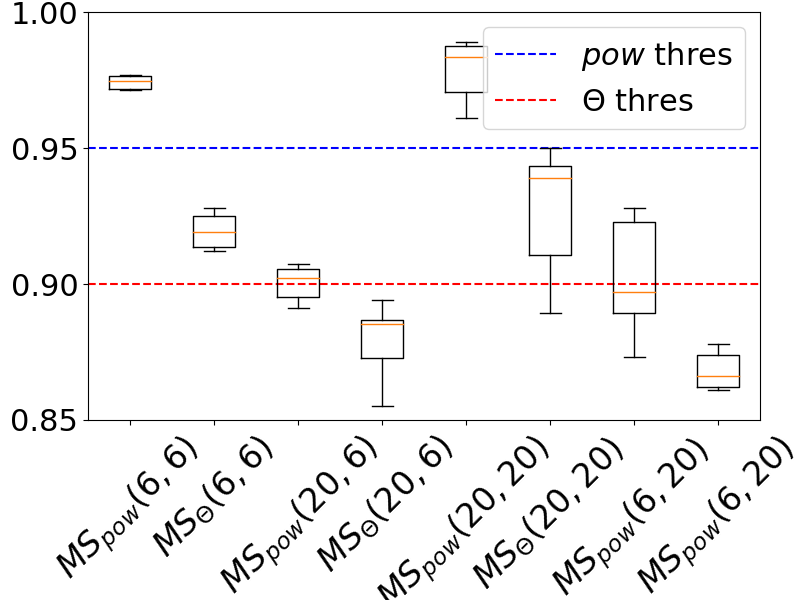}
		\caption{MSs of two most similar sensors}
		{\footnotesize Note: $MS(i, j)$ means the $MS$ computed by testing the RF signals from device No. i by the model No. j \par}
		\label{fig:specialSensors}
	\end{minipage}%
	\hspace{\fill}
	\begin{minipage}[t]{0.48\columnwidth}
		\centering
		\includegraphics[width=1.0\columnwidth]{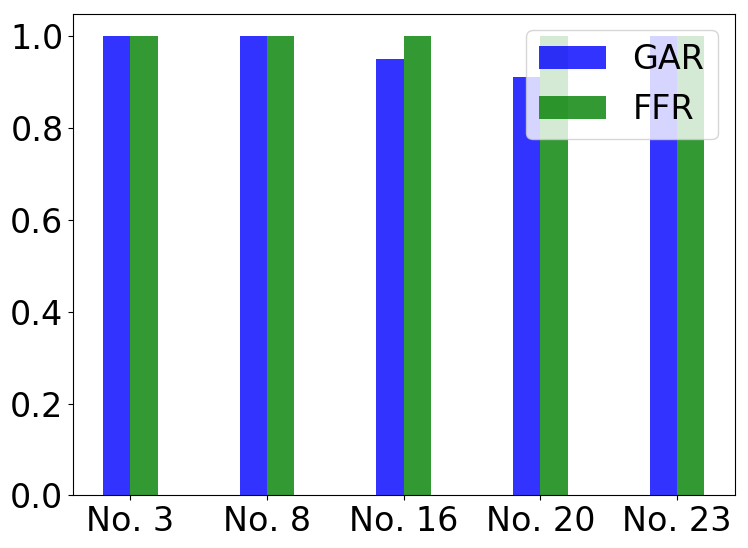}
		\caption{GARs and FFRs for $5$ worst devices in multipath scenarios after implementing the workarounds}
		\label{fig:bad_device}
	\end{minipage}
	
\end{figure}

For all the tested devices, regardless of the communication distance and orientation, the GAR and FRR can be both $100\%$ and the GRR and FAR can be both $0\%$ by testing two consecutive packets each time. Moreover, two pairs of sensors that are indistinguishable by the modulation error-based system can be recognized by $\mathcal{F}$ID accurately. As shown in Fig. \ref{fig:specialSensors}, two devices are indeed very similar, since the $MS_{pow}$s are $0.87 \sim 0.92$ and the $MS_{\Theta}$s are $0.86 \sim 0.89$ when testing the RF signals from one device by the other device's function model. However, $\mathcal{F}$ID can still identify them accurately, as shown in table \ref{tab:performance}.

\subsection{Robustness Analysis}
\label{sec:multipath}

We mainly consider $2$ scenarios: spatial variation scenario and multipath scenario. In spatial variation scenario, as stated in section \ref{sec:lineofsight}, we conduct experiments in line-of-sight environments where the communication distance and orientation can vary. The experiment results indicate that $\mathcal{F}$ID is robust to these two significant spatial variations.

In the multipath scenario, we arbitrarily place $2$ or $3$ metal boxes between the transmitter and the receiver to create multipath channels.
When testing $\mathcal{F}$ID on wireless sensor nodes working with the $IEEE 802.15.4$ protocol, we found that $\mathcal{F}$ID can still work well even without channel estimation and deconvolution. The reason is that the symbol rate for $IEEE 802.15.4$ is only $1M/s$, and hence in one symbol period, the RF signal can travel for $300m$. However, in reality, the distinctions between different paths are approximately tens of meters. So the RF signal coming from the line-of-sight path is very similar to the signals from the other paths. Therefore, the combination of those signals is similar to multiplying the RF signal from the line-of-sight path by a factor. And since the nonlinear terms are extracted from normalized RF signals, this factor does not affect the identification results. However, if we want to identify wifi signals whose symbol rate is 20M/s in multipath environments, channel estimation and deconvolution is an indispensable step, which should be added into the first module of $\mathcal{F}$ID.
Besides, considering that the multipath fading might attenuate the RF signal strength severely, before identifying the received RF signal, we need to confirm that the SNR should be at least $10dB$ and make the decision based on two adjacent packets as mentioned before. Since the channel situation is more complicated, the $MS_{pow}$ and $MS_{\Theta}$ can not be that high as in the line-of-sight environments. Therefore, we need to adjust the predefined thresholds. Specifically, for $MS_{pow}$, the threshold is reduced to $0.9$, and for $MS_{\Theta}$, the threshold is reduced to $0.85$ for some devices and remains $0.9$ for the others based on the previous modeling results.
After all these modifications, the GAR and FRR remain $1.0$ for most of the tested devices except Device No.3, No. 8, No. 16, No. 20, and No. 23. For these $5$ devices, the $pow^r(x_n(t), \tau)$ can vary a lot even in one symbol period, and hence the combination of the signals from different paths can not be simplified as the product of the RF signal from the line-of-sight path and a factor. Therefore, using the aforementioned simplification assumption will degrade $MS_{pow}$ and thus the GARs for those $5$ devices in this case. To alleviate this problem, we implement two workarounds for those $5$ devices: 1. Implement channel estimation and deconvolution and identify the deconvoluted $z[n]$. 2. Identify the RF signal only by carrier frequency offset and $MS_{\Theta}$. Fig. \ref{fig:bad_device} shows that the GARs for those $5$ devices are improved to over $0.9$ so that the overall $BIA$ in multipath environments is over $99\%$.

\section{Conclusion and Future Work}
RF fingerprinting is a cost-efficient identification method for low-end IoT devices. In this paper, we propose a function model, which is a high-precision approximation of the mathematical expression of the physical-layer process from modulation to power amplification, as RF fingerprint for device identification. A data-independent and channel-robust RF fingerprinting system is further designed based on our function model, namely $\mathcal{F}$ID. $\mathcal{F}$ID is evaluted in various scenarios, and it achieves over $99\%$ accuracy overall.

$\mathcal{F}$ID is a successful trial to break the convention of designing a feature-based RF fingerprinting system. Our basic idea is straightforward since similar approaches have been used for identifying power amplifiers \cite{Adam2011}. However, we are the first to demonstrate that such kind of methods can work on real devices, and our function model is also a novel accurate and efficient model. These are our two main technical contributions in this paper.
However, there are several open issues related to $\mathcal{F}$ID that require further study. For instance, $\mathcal{F}$ID will bring a new challenge for the security research in this area, because $\mathcal{F}$ID can reproduce almost the same RF signals as the authenticated devices produce. Specifically, by using $\mathcal{F}$ID, the attackers can simply compute the transmitted signal by the references of the linear parts and the nonlinear terms by the pre-trained KR models. If the attackers use a high-end RF transceiver, the reproduced signals will be very similar to the genuine signals. Therefore, our function modeling method can be applied to attack most existing RF fingerprinting systems, and it is challenging to defend this attack.

\textbf{Acknowledgement:} This research is mainly supported by National Science Foundation under Grant No. 1421903.

\end{document}